\documentclass[twocolumn]{aastex62} %linenumbers
\usepackage{amsmath}
\usepackage{longtable, booktabs, threeparttablex}
\graphicspath{{./}{figures/}}

\shorttitle{Orbit of a Possible Planet X}
\shortauthors{Siraj, Chyba, \& Tremaine}
\begin{document}

\title{Orbit of a Possible Planet X}

\email{siraj@princeton.edu}

\author{Amir Siraj}
\affil{Department of Astrophysical Sciences, Princeton University, 4 Ivy Lane, Princeton, NJ 08544, USA}

\author{Christopher F. Chyba}
\affil{Department of Astrophysical Sciences, Princeton University, 4 Ivy Lane, Princeton, NJ 08544, USA}
\affil{School of Public and International Affairs, Princeton University, 20 Prospect Lane, Princeton, NJ 08540, USA}

\author{Scott Tremaine}
\affil{School of Natural Sciences, Institute for Advanced Study, Princeton, NJ 08540, USA}

\begin{abstract}

The plausibility of an unseen planet in the outer solar system, and the expected orbit and mass of such a planet, have long been a topic of inquiry and debate. We calculate the long-term orbital stability of distant TNOs, which allows us to expand the sample of objects that would carry dynamical information about a hypothetical unseen planet in the solar system. Using this expanded sample, we find statistically significant clustering at the $\sim 3 \sigma$ level for TNOs with semimajor axes $>170\mbox{\;AU}$, in longitude of perihelion ($\varpi$), but not in inclination ($i$), argument of perihelion ($\omega$) or longitude of node ($\Omega$). Since a natural explanation for clustering in $\varpi$ is an unseen planet, we run 300 $n$-body simulations with the giant planets, a disk of test particles representing Kuiper belt objects, and an additional planet with varied initial conditions for its mass, semimajor axis, eccentricity, and inclination. Based on the distribution of test particles after 1--2 Gyr, we compute relative likelihoods given the actual distribution of $\varpi$ as a function of semimajor axis for distant TNOs on stable orbits using a significantly larger sample than previous work. We find the best-fit unseen planet parameters to be: mass $m_p = 4.4\pm1.1\mathrm{\;M_{\oplus}}$, semimajor axis $a_p=290\pm30\mathrm{\;AU}$, eccentricity $e_p=0.29\pm0.13$, and inclination $i_p=6.8\pm5.0^{\circ}$. Only $0.06\%$ of the \cite{2021AJ....162..219B} Planet Nine reference population produce probabilities within $1\sigma$ of the maximum within our quadrivariate model, indicating that our work identifies a distinct preferred region of parameter space for an unseen planet in the solar system.

\end{abstract}


\section{Introduction}

There is a long history of theoretically proposed planets in the outer solar system, dating back to the mid-1800s \citep{9546ef17-23ca-3ae5-bacb-94746b4a39b5,standish1993,fienga2020}. Recently, the structure of the distant Kuiper belt has led to speculation regarding the possibility of an unseen planet \citep{2006ApJ...643L.135G, 2008AJ....135.1161L, 2014Natur.507..471T, 2016AJ....151...22B, 2017AJ....154...62V, 2018AJ....155...75S, 2022ApJ...938L..23H, 2023AJ....166..118L}. Some of these recent studies have been motivated by apparent clustering of distant trans-Neptunian objects (TNOs) in various orbital parameters, including longitude of perihelion ($\varpi$), longitude of the ascending node ($\Omega$), argument of perihelion ($\omega\equiv\varpi-\Omega$) and inclination relative to the ecliptic ($i$) \citep{2014Natur.507..471T, 2016AJ....151...22B, 2016AJ....152..221S}. An unseen planet in the outer solar system (sometimes referred to as ``Planet X'' or ``Planet Nine'') could potentially shepherd the orbits of distant TNOs into clustered configurations \citep{2016AJ....151...22B}. The observational search for such a planet has, to date, been unsuccessful \citep{2024AJ....167..146B}.

There is an ongoing debate over whether the claimed clustering of distant TNOs is real or spurious, perhaps arising from observational selection effects or limited statistics. \cite{2017AJ....154...50S} and \cite{2020PSJ.....1...28B} could not conclude that distant TNOs were clustered in Dark Energy Survey (DES) data alone, and \cite{2022ApJS..258...41B} similarly could not conclude that distant TNOs were clustered in Outer Solar System Origins Survey (OSSOS) data alone. \cite{2019AJ....157...62B} and \cite{2021PSJ.....2...59N} examined larger samples of TNOs and reached opposite conclusions about the statistical significance of clustering in the orbital elements.\footnote{Note that \cite{2021PSJ.....2...59N} do not exclude clustering at the level claimed by \cite{2019AJ....157...62B} but rather find that the observed distributions of various angular orbital elements for a sample of distant TNOs within surveys with well-characterized footprints are consistent with uniform distributions.} In addition to clustering in angular orbital elements, an unseen planet could also produce a population of high-inclination Centaurs \citep{2016ApJ...833L...3B, 2019PhR...805....1B, 2019AJ....158...43K}.

\cite{2021AJ....162..219B} ran a suite of 121 $n$-body simulations testing various parameters for an additional planet in the solar system. Each simulation contained several tens of thousands of test particles, whose orbital parameters were compared to the $\varpi$, $\Omega$, and $i$ distributions of 11 distant TNOs to  identify a preferred region of parameter space for the additional planet. Additionally, \cite{2024ApJ...966L...8B} argued that an unseen planet may produce a population of Neptune-crossing TNOs significantly more consistent with the observed population of such objects than if no unseen planet were present. 

In this paper, we re-examine the question of whether or not the current distribution of distant TNO detections suggests clustering in $\varpi$, $\Omega$, and $i$. Furthermore, we explore the parameter space of hypothetical unseen planets and ask what parameters are  most likely given the current state of observations. 

A novel feature of this work is that we determine the long-term stability for a large set of distant TNOs -- 
such stability is crucial for evaluating the plausibility of an unseen planet because it takes $\sim 1 \mathrm{\; Gyr}$ for such a planet to induce clustering amongst TNOs \citep{2021AJ....162..219B}.
Using this information and new discoveries of distant TNOs in addition to a broader range of allowed semimajor axes we are able to expand the sample of TNOs that we search for clustering to 51 objects.

\begin{figure}%[hptb]
 \centering
\includegraphics[width=\linewidth]{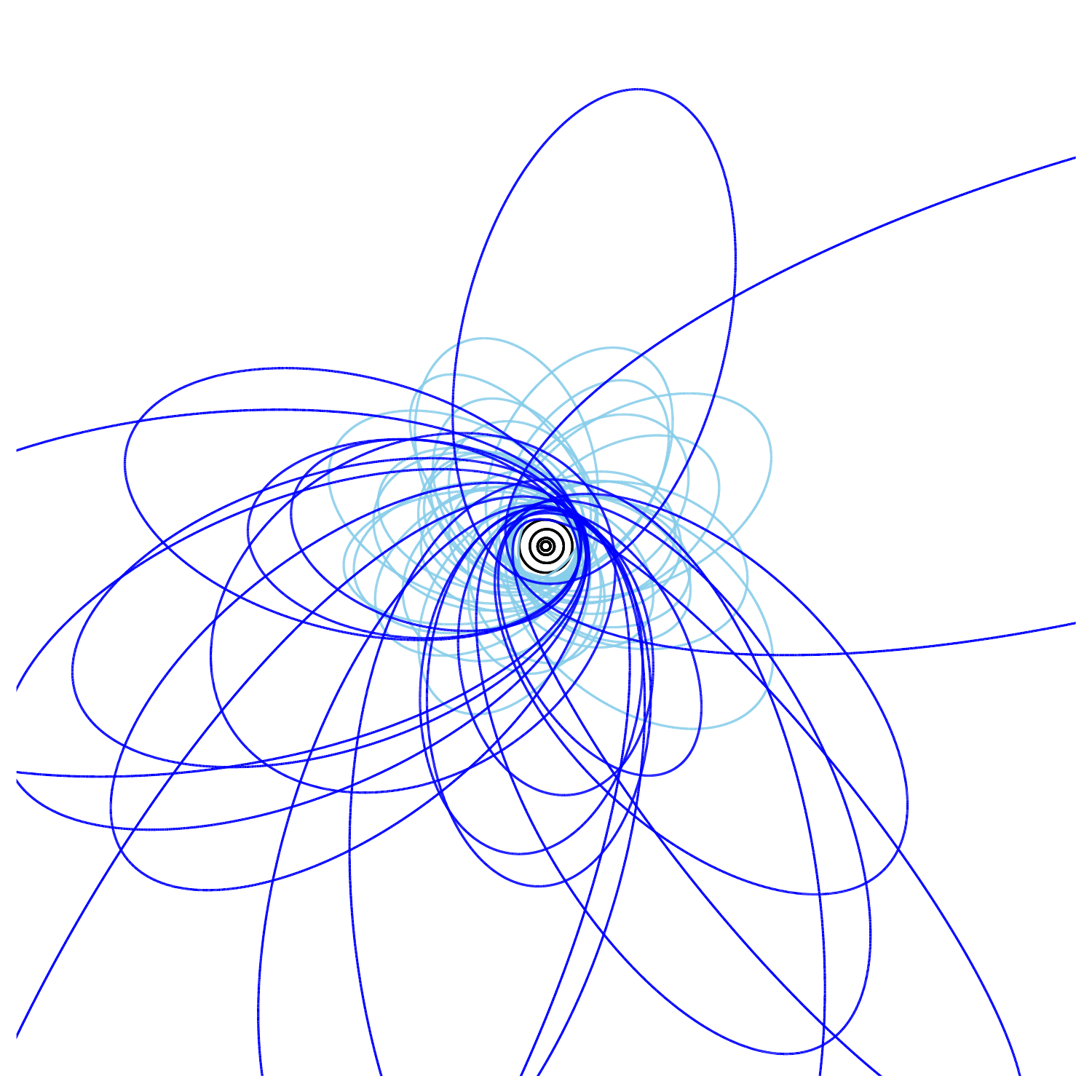}
\caption{Orbits of objects listed in Table \ref{stableobjects}, which are all stable or metastable. Orbits with $a > 170 \mathrm{\; AU}$ are illustrated in dark blue while orbits with $a < 170 \mathrm{\; AU}$ are in light blue. Clustering of the directions of the aphelia of the orbits is apparent for all but two of the distant (dark blue) objects. The black orbits at the center are those of the giant planets. Overall scale is $600 \mathrm{\; AU} \times 600 \mathrm{\; AU}$.}
\label{fig:orbitdiagram}
\end{figure}

\section{Clustering Analysis}
\label{clusteringanalysis}

\subsection{Long-term stability}
\label{longtermstability}

The orbital shepherding mechanism observed in \textit{n}-body simulations of the solar system plus an additional trans-Neptunian planet and a distribution of test particles occurs on a timescale of order $\sim 1 \mathrm{\; Gyr}$ \citep{2021AJ....162..219B}. As a result, clustering analyses that test the effects of this mechanism should be limited to examining the subset of trans-Neptunian objects that remain stable to perturbations from the solar system's giant planets (Neptune, Uranus, Saturn, and Jupiter) on Gyr timescales. While a cutoff of around $q > 42 \mathrm{\; AU}$ has been used in the past as a heuristic criterion for stability \citep{2008ssbn.book..259G, 2015MNRAS.446.3788B, 2021AJ....162..219B}, the best way to check stability of any particular object is to do so directly with long-term $n$-body simulations.

We adopt the definitions of \cite{2021ApJ...910L..20B} for unstable, metastable, and stable objects, as follows. For each object we create a set of $200$ clones, by selecting a value for each of the six Keplerian orbital elements from the respective Gaussian distribution defined by the nominal value and the associated standard deviation as reported by the JPL Small-Body Database\footnote{\url{https://ssd.jpl.nasa.gov/tools/sbdb_query.html}}. We then integrate the orbit of each clone for $4$ Gyr. 
Unstable objects are those for which $\ge 20 \%$ of clones become unbound from the Sun, metastable objects are those that are not unstable but for which $\ge 20 \%$ of clones exhibit a change in semimajor axis by a factor of $\ge 2$, and stable objects are those that are neither unstable nor metastable. We carry out all $n$-body simulations in this work using the \texttt{REBOUND} package \citep{2012A&A...537A.128R} with the \texttt{MERCURIUS} hybrid symplectic integrator \citep{2019MNRAS.485.5490R, 2024MNRAS.533.3708L}, using a timestep of 0.6 yr, which satisfies the \cite{2015AJ....150..127W} pericenter resolution criterion for all outer solar system orbits that do not interact strongly with Jupiter. This choice is suitable for the purposes of this work, since the focus is on objects with perihelion $q \gtrsim 30 \mathrm{\; AU}$, far larger than Jupiter's semimajor axis of $5.2\mathrm{\; AU}$.

Table \ref{stableobjects} lists orbital parameters of the 51 stable or metastable TNOs with $q>30\mathrm{\,AU}$ and $90\mbox{\,AU} < a < 1000 \,\mbox{AU}$,\footnote{All orbital elements are heliocentric, although we note that all objects in Table 1 fall within the range $90\mbox{\,AU} - 1000 \,\mbox{AU}$ for both heliocentric and barycentric $a$, and within the range $\sim 35$--$80 \mathrm{\; AU}$ for both heliocentric and barycentric $q$.} as well as the percentage of all clones that survive (those that do not become unbound from the Sun or collide with the Sun or a planet) and the percentage of all clones that undergo a change in semimajor axis of a factor of two or more (excluding those that do not survive). The orbits of the objects listed in Table \ref{stableobjects} are illustrated in Figure \ref{fig:orbitdiagram}. In addition, there are 96 unstable TNOs in the same range of perihelion and semimajor axis. Of the 51 stable/metastable TNOs, 30 have perihelion distance in the range $35.4\mbox{\;AU} \leq q \leq 42\mbox{\;AU}$ that lies inside the traditional `stability boundary' for TNOs on highly eccentric orbits.

The 11 objects included by \cite{2021AJ....162..219B} in their latest comparison of simulations of the solar system including a hypothetical trans-Neptunian planet with observations of distant TNOs are marked with asterisks in Table \ref{stableobjects}. This sample consisted of all TNOs with $q>42\mathrm{\,AU}$ and $150\mbox{\,AU} < a < 1000 \,\mbox{AU}$ at the time of writing. We find that there are 24 stable or metastable objects over the same range of semimajor axis. This increase is due in part to newly discovered objects and improved orbits for existing objects since 2021 and in part due to stable/metastable objects with $q<42\mathrm{\,AU}$. 

Throughout this paper, our ``fiducial sample'' refers to all of the objects in Table \ref{stableobjects} plus all of the unstable objects in the same range of perihelion and semimajor axis, $q>30\mathrm{\,AU}$ and $90\mbox{\,AU} < a < 1000 \,\mbox{AU}$. Note that we exclude any TNOs with an excessively large uncertainty parameter\footnote{See \url{http://www.minorplanetcenter.org/iau/info/UValue.html}. The MPC ``uncertainty parameter'' is denoted the ``condition code'' in the JPL database.} ($> 6$) at the time of writing. 

We note that the choice of a lower cutoff for semimajor axis represents a tension between the reduction of Neptune's gravitational influence at larger semimajor axes and the fact that the number of discovered objects falls off steeply as a function of semimajor axis. For example, if a population of $\gtrsim 10^2$ objects are required within the range $a_{\rm min} < a < 170\mathrm{\;AU}$, one can choose $a_{\rm min} = 90 \mathrm{\; AU}$, which is beyond the $n:1$ Neptune mean-motion-resonances for $n \leq 5$. We find that none of our results (namely the Kuiper test for $\varpi$ described in Section \ref{kuipertestsection} and the bias analyses for $\varpi$ and $\Omega$ described in Section \ref{bias}) decrease in statistical significance when $a_{\rm min}$ is perturbed by $\pm 5 \mathrm{\; AU}$ from our nominal minimum value of $90 \mathrm{\; AU}$.

%\newpage
\subsection{Kuiper test}
\label{kuipertestsection}

\begin{figure}%[hptb]
 \centering
\includegraphics[width=\linewidth]{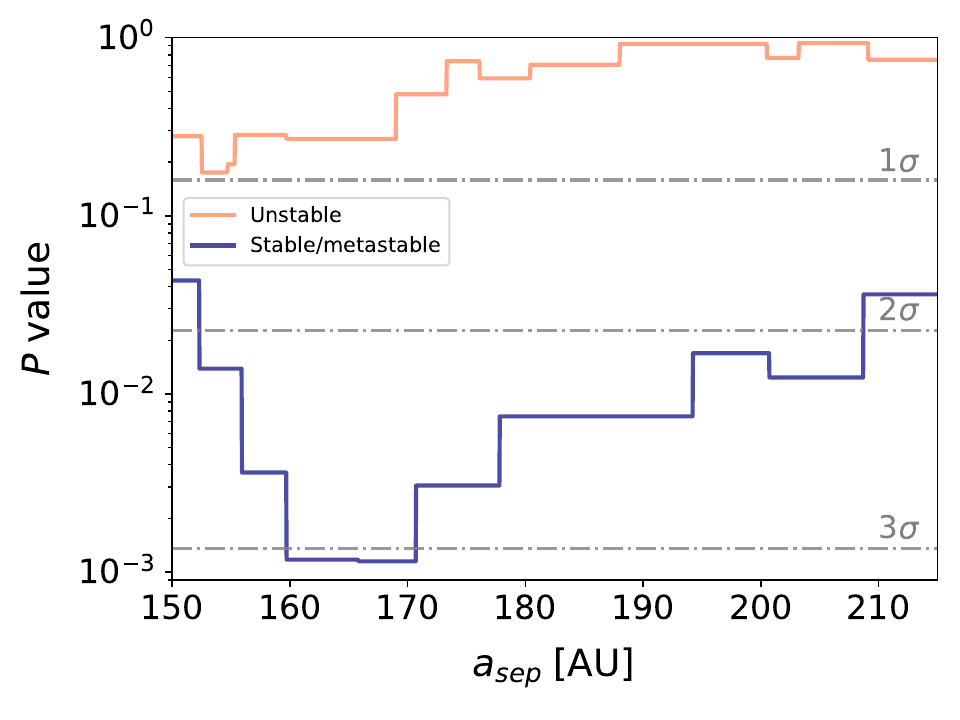}
\caption{Two-sample Kuiper test comparing the distribution in $\varpi$ as a function of $a_{\rm sep}$ between subsets $90 \mathrm{\; AU} < a < a_{\rm sep}$ and $a_{\rm sep} < a < 10^3 \mathrm{\; AU}$ of all TNOs with $q > 30 \mathrm{\; AU}$. Red corresponds to the unstable population, while blue corresponds to the stable/metastable population. The $y$-axis indicates the probability the subsets on each side of $a_{\rm sep}$ were drawn from the same distribution.}
\label{fig:kuiper_statistic}
\end{figure}

\begin{figure}%[hptb]
 \centering
\includegraphics[width=\linewidth]{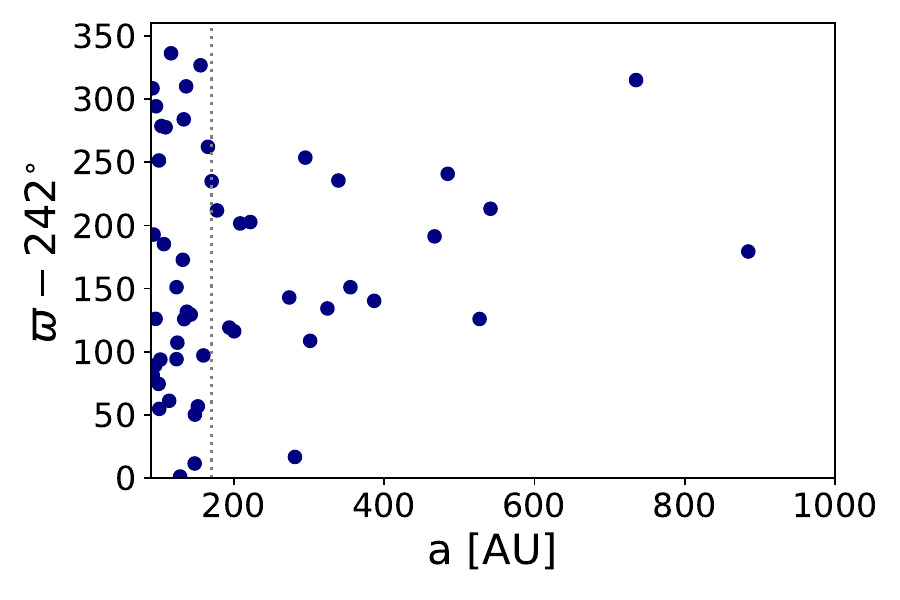}
\caption{Distribution of $\varpi$ for all objects in Table \ref{stableobjects}. Line at $a = a_{\rm sep}=170 \mathrm{\; AU}$ for reference. Objects with $a>a_{\rm sep}$ exhibit apsidal clustering.}
\label{fig:tno_pomegadist}
\end{figure}

Simulations of the outer solar system with an additional outer planet often produce apsidal clustering in $\varpi$ with a sharp onset as a function of increasing semimajor axis \citep{2021AJ....162..219B}. If such an architecture exists in the solar system and is caused by an unseen planet, it should be apparent in the set of stable and metastable TNOs since they have lifetimes long enough ($\gtrsim 1 \mathrm{\; Gyr}$) to be gravitationally shepherded by an unseen planet, but absent or nearly so in the set of unstable TNOs.

We use the Kuiper test, a rotationally invariant version of the Kolmogorov--Smirnov test, to search for a sharp transition in the $\varpi$ distribution as a function of semimajor axis $a$. Specifically, we evaluate the likelihood $p$ that the distribution of $\varpi$ for the subset of TNOs interior to a given separating value of semimajor axis $a_{\rm sep}$ could have been drawn from the same parent distribution as the TNOs exterior to $a_{\rm sep}$. We evaluate the $p$ values given by the Kuiper test as a function of $a_{\rm sep}$, for both the stable/metastable and unstable populations.  

The results are displayed in Figure \ref{fig:kuiper_statistic}. For the stable/metastable population, there is a $p
= 10^{-3}$ or $>3\sigma$ transition in the distribution of $\varpi$ at $a \sim 170 \mathrm{\; AU}$. In contrast, the unstable population shows no evidence of a transition, with $p>0.2$ ($<1 \sigma$) across a wide range of separating semimajor axes ($a_{\rm sep}$). Note that the $p$ value increases as $a_{\rm sep}$ gets further from $\sim 170 \mathrm{\; AU}$ because the unclustered and clustered populations become mixed (within the inner sample for $a_{\rm sep} \gtrsim 170 \mathrm{\; AU}$, and within the outer sample for $a_{\rm sep} \lesssim 170 \mathrm{\; AU}$).

Could the difference between the angular distributions of the unstable and stable/metastable populations be rooted in an observational selection effect? Over the past $1 \mathrm{\; Gyr}$, following Neptune's orbit under the gravitational influence of Uranus, Saturn, Jupiter, and the Sun, we find that Neptune's maximum eccentricity and inclination were $0.016$ and $2.4^{\circ}$, respectively. Given these low values, the shape and orientation of Neptune's orbit cannot cause a strong angular dependency in the stability of TNOs. We also check whether a selection effect on brightness or on-sky rate of motion could explain the difference between the unstable and stable/metastable populations. We consider objects within the fiducial sample for which the discovery circumstances are documented in the JPL Small-Body Database\footnote{\url{https://ssd.jpl.nasa.gov/discovery.html}} ($\sim 30\%$ across both the unstable the stable/metastable subsets), and perform a two-sided Kolmogorov–Smirnov test (unstable versus stable/metastable) on apparent magnitude at discovery and on apparent rate of motion at discovery. We find statistically insignificant results for both tests: $p = 0.4$ and $p = 0.2$, respectively, indicating a lack of evidence for such observational selection effects. The mean and standard deviation of the distribution of apparent magnitude at discovery for the unstable and stable/metastable subsets are $22.2 \pm 0.8$ and $22.3 \pm 1.1$, respectively. The mean and standard deviation of the distribution of apparent motion at discovery for the unstable and stable/metastable subsets are $2.5 \pm 0.9 \mathrm{\; arcsec/hr}$ and $2.3 \pm 0.8 \mathrm{\; arcsec/hr}$, respectively.

When the same tests are performed for the longitude of node $\Omega$, the argument of perihelion $\omega$, and the inclination $i$ -- with $a_{\rm sep}$ as a free parameter for each -- they yield no statistically significant differences between the unstable and the stable/metastable populations of our fiducial sample.\footnote{For $a_{\rm sep} = 170 \mathrm{\; AU}$, Kuiper tests on the stable/metastable and unstable populations yielded $p = 0.76$ and $p = 0.77$ for longitude of node, $p = 0.13$ and $p = 0.33$ for inclination, and $p = 0.56$ and $p = 0.25$ for argument of perihelion, respectively.} As a result, we focus exclusively on $\varpi$ clustering when comparing the output of $n$-body simulations to the distribution of known TNOs later in this paper (Section \ref{sec:simul}). While evidence for clustering in $\Omega$ and $i$ is not required for consistency with an unseen planet, such clustering has been invoked in previous work \citep{2021AJ....162..219B}. 

Figure \ref{fig:tno_pomegadist} shows the distribution of the longitude of perihelion $\varpi$ for all the objects in Table \ref{stableobjects}, providing visual confirmation of the apsidal clustering for stable/metastable TNOs with $a>a_{\rm sep}=170\mbox{\;AU}$. We perform an additional test in Section \ref{evector} without cuts on discovery survey, as well as a suite of tests across the subset of the fiducial sample discovered by surveys with well-characterized selection functions in Sections \ref{bias} and \ref{simbias}, which confirm the conclusions drawn here: statistically significant clustering is present in the longitude of perihelion $\varpi$ for semimajor axes $a \gtrsim 170\mbox{\;AU}$, but not in other angular orbital elements.

\subsection{Eccentricity vector}
\label{evector}

The eccentricity vector is defined as
\begin{equation}
    \mathbf{e} = \frac{\dot{\mathbf{r}} \times (\mathbf{r} \times \dot{\mathbf{r}})}{GM} - \frac{\mathbf{r}}{r} \; \; ,
\end{equation}
its magnitude equals the eccentricity and it points towards perihelion \citep{2023dyps.book.....T}.

For any population of TNOs, the mean eccentricity vector can be computed. 
To test the significance of any asymmetry in the orientation of the orbits in 3D space, we compare the magnitude of the mean eccentricity vector to the distribution of mean eccentricity vector magnitudes if the orbits were randomly oriented. To do so, we initialize a large number of copies ($10^6$) of the TNO population, and in each copy we replace the values of $\varpi$ and $\Omega$ with randomly drawn angles in the range $0$--$2 \pi$. We then compute the mean eccentricity vector for each randomly oriented copy of the original population. Finally, we ask what fraction of the randomly oriented copies have a mean eccentricity vector magnitude exceeding that of the actual population.

We run this test separately for the stable/metastable and unstable subsets of our fiducial sample with $a > 170 \mathrm{\; AU}$. We find that the fraction of randomly oriented copies that exceed the magnitude of the actual mean eccentricity vector for the stable/metastable subset ($\langle\mathbf{e}\rangle=0.44$) is only $p\sim 2 \times 10^{-3}$. In contrast, the fraction of randomly oriented copies that exceed the magnitude of the actual mean eccentricity vector for the unstable subset ($\langle\mathbf{e}\rangle=0.26$) is $p\sim 0.2$. The two-order-of-magnitude difference in $p$ between the stable/metastable and unstable subsets argues against selection effects as the source of the asymmetry.

\begin{ThreePartTable}

\begin{TableNotes}
  \item[] \textsc{Note --} Asterisks mark the 11 objects in the \cite{2021AJ....162..219B} sample. [M] denotes metastable objects. The angles $\varpi$, $\Omega$ and $i$ are in ecliptic coordinates. All orbital elements are heliocentric and provided by the JPL Small-Body Database.
\end{TableNotes}

\begin{longtable*}{@{\extracolsep{\fill}}c c c c c c c c@{}}
\caption{Stable/metastable objects with $q > 30 \mathrm{\; AU}$ and $90 \mathrm{\; AU}< a < 1000 \mathrm{\; AU}$.}\\
\hline
\multicolumn{1}{c}{Name} & \multicolumn{1}{c}{$a$ [AU]} & \multicolumn{1}{c}{$q$ [AU]} &  \multicolumn{1}{c}{$i$ [deg]} & \multicolumn{1}{c}{$\varpi$ [deg]} & \multicolumn{1}{c}{$\Omega$ [deg]} & \multicolumn{1}{c}{Survival \%} & \multicolumn{1}{c}{Shift \%}\\
\hline
\hline
1996 GQ21   & 92.1       & 38.2       & 13.38       & 190.46                & 194.32                & 99          & 3          \\
2013 RG124  & 92.3       & 39         & 43.46       & 323.52                & 35.7                  & 81          & 3          \\
2014 FJ72   & 93.5       & 38.1       & 15.38       & 74.71                 & 302.71                & 93          & 6          \\
2005 RP43   & 95.8       & 38.4       & 22.89       & 331.27                & 344.16                & 83          & 6          \\
2014 QR562  & 96.2       & 39.9       & 28.26       & 8.02                  & 148.17                & 100         & 2          \\
2010 ER65   & 96.7       & 40         & 21.31       & 176.23                & 212.61                & 100         & 2          \\
2000 OM67   & 100.2      & 39.3       & 23.33       & 316.53                & 327.04                & 96          & 4          \\
2014 US277  & 100.6      & 54.9       & 36.36       & 133.34                & 143.2                 & 100         & 0          \\
2008 ST291  & 101        & 42.7       & 20.75       & 296.77                & 330.95                & 100         & 0          \\
2013 VS46   & 102.5      & 37.2       & 20.17       & 335.8                 & 69.4                  & 80          & 11         \\
2017 FE173  & 103.9      & 42         & 31.03       & 160.61                & 178.97                & 100         & 0          \\
2014 ST373  & 107.2      & 50.2       & 43.15       & 67.16                 & 130.33                & 100         & 0          \\
2014 UZ224  & 109.5      & 38.8       & 26.78       & 159.61                & 130.91                & 86          & 7          \\
2005 QU182  & 114.2      & 37.1       & 14.01       & 303.17                & 78.51                 & 87          & 14         \\
2020 BB95   & 116.7      & 38.4       & 18.18       & 218.17                & 91.65                 & 90          & 7          \\
2014 RV86   & 124        & 39.3       & 28.49       & 33.07                 & 129.57                & 99          & 3          \\
2013 RE124  & 124        & 40.1       & 31.61       & 336.11                & 175.64                & 95          & 5          \\
2014 MJ70   & 125.1      & 37.7       & 15.52       & 349.16                & 96.04                 & 90          & 16         \\
2015 KE172  & 128.8      & 44.2       & 38.44       & 243.22                & 227.64                & 100         & 0          \\
2014 QS562  & 132.3      & 40.4       & 24.91       & 54.76                 & 77.32                 & 98          & 3          \\
2022 GV6    & 133.6      & 38.2       & 13.71       & 165.89                & 200.68                & 91          & 15         \\
2007 TC434  & 134.2      & 39.6       & 26.41       & 7.79                  & 16.19                 & 80          & 14         \\
2014 JW80   & 136.8      & 38.1       & 40.89       & 192                   & 60.8                  & 84          & 12         \\
2014 VP43   & 137.7      & 41         & 35.82       & 13.67                 & 59.35                 & 97          & 5          \\
2014 SQ403  & 142.9      & 46.6       & 43.99       & 11.41                 & 180.18                & 98          & 1          \\
2013 GP136  & 148.1      & 41         & 33.62       & 253.55                & 210.78                & 91          & 4          \\
2014 SS349  & 148.4      & 45.5       & 48.18       & 292.25                & 144.16                & 97          & 1          \\
2015 KH163  & 152.3      & 39.9       & 27.19       & 298.75                & 67.58                 & 91          & 7          \\
2003 HB57   & 155.9      & 38.1       & 15.53       & 208.67                & 197.96                & 82          & 26 {[}M{]} \\
2005 RH52   & 159.7      & 39.1       & 20.44       & 339.05                & 306.09                & 89          & 18         \\
2021 LP43   & 165.8      & 39.5       & 9           & 144.18                & 134.24                & 96          & 24 {[}M{]} \\
2016 TP120  & 170.7      & 40.4       & 32.64       & 116.9                 & 126.69                & 82          & 11         \\
2016 QV89   & 177.8      & 40.1       & 21.36       & 93.81                 & 173.1                 & 98          & 10         \\
2015 UN105  & 194.2      & 41.4       & 37.02       & 1.11                  & 129.32                & 98          & 9          \\
2003 SS422  & 200.7      & 39.6       & 16.78       & 358.14                & 150.99                & 88          & 34 {[}M{]} \\
2013 UT15*  & 208.7      & 44.1       & 10.62       & 83.5                  & 191.91                & 100         & 0          \\
2000 CR105* & 222.1      & 43.9       & 22.77       & 84.62                 & 128.3                 & 100         & 2          \\
2012 VP113* & 273.9      & 80.6       & 24.02       & 24.92                 & 90.85                 & 100         & 0          \\
2013 FT28*  & 281.5      & 43.4       & 17.43       & 258.71                & 217.69                & 100         & 17         \\
2018 VM35   & 295.3      & 44.7       & 8.47        & 135.6                 & 192.23                & 87          & 13         \\
2014 WB556  & 301.7      & 42.8       & 24.15       & 350.59                & 114.89                & 100         & 16         \\
2014 SR349* & 324.7      & 47.5       & 17.94       & 16.23                 & 35.02                 & 100         & 1          \\
2010 GB174* & 339.3      & 48.4       & 21.57       & 117.49                & 130.77                & 100         & 1          \\
2004 VN112* & 355.3      & 47.3       & 25.5        & 33.05                 & 66.04                 & 100         & 4          \\
2016 SD106  & 386.9      & 42.7       & 4.8         & 22.26                 & 219.5                 & 82          & 41 {[}M{]} \\
2015 RX245* & 467.2      & 45.7       & 12.1        & 73.26                 & 8.58                  & 100         & 33 {[}M{]} \\
2015 BP519  & 484.7      & 35.4       & 54.12       & 122.77                & 135.15                & 84          & 42 {[}M{]} \\
2013 RA109* & 527.1      & 46         & 12.4        & 7.91                  & 104.82                & 100         & 39 {[}M{]} \\
2003 VB12*  & 541.6      & 76.4       & 11.93       & 95.14                 & 144.3                 & 100         & 0          \\
2021 DK18   & 735.2      & 44.5       & 15.45       & 196.98                & 322.22                & 82          & 26 {[}M{]} \\
2013 SY99*  & 884.5      & 50.1       & 4.21        & 61.31                 & 29.53                 & 100         & 23 {[}M{]} \\ 
\hline
\insertTableNotes
\label{stableobjects}
\end{longtable*}
%\end{table*}
\end{ThreePartTable}
%\newpage
\section{Simulations}
\label{sec:sim}

Given the statistically significant clustering in $\varpi$, we are motivated to explore how various models for an additional planet in the solar system would shape the $\varpi$ distribution of distant TNOs. To do so, we run a suite of numerical $n$-body simulations including the Sun, the four giant planets, a disk of massless test particles, and an additional planet exterior to Neptune. 
\label{sec:simul}

\begin{figure*}%[hptb]
\centering
\includegraphics[width=\linewidth]{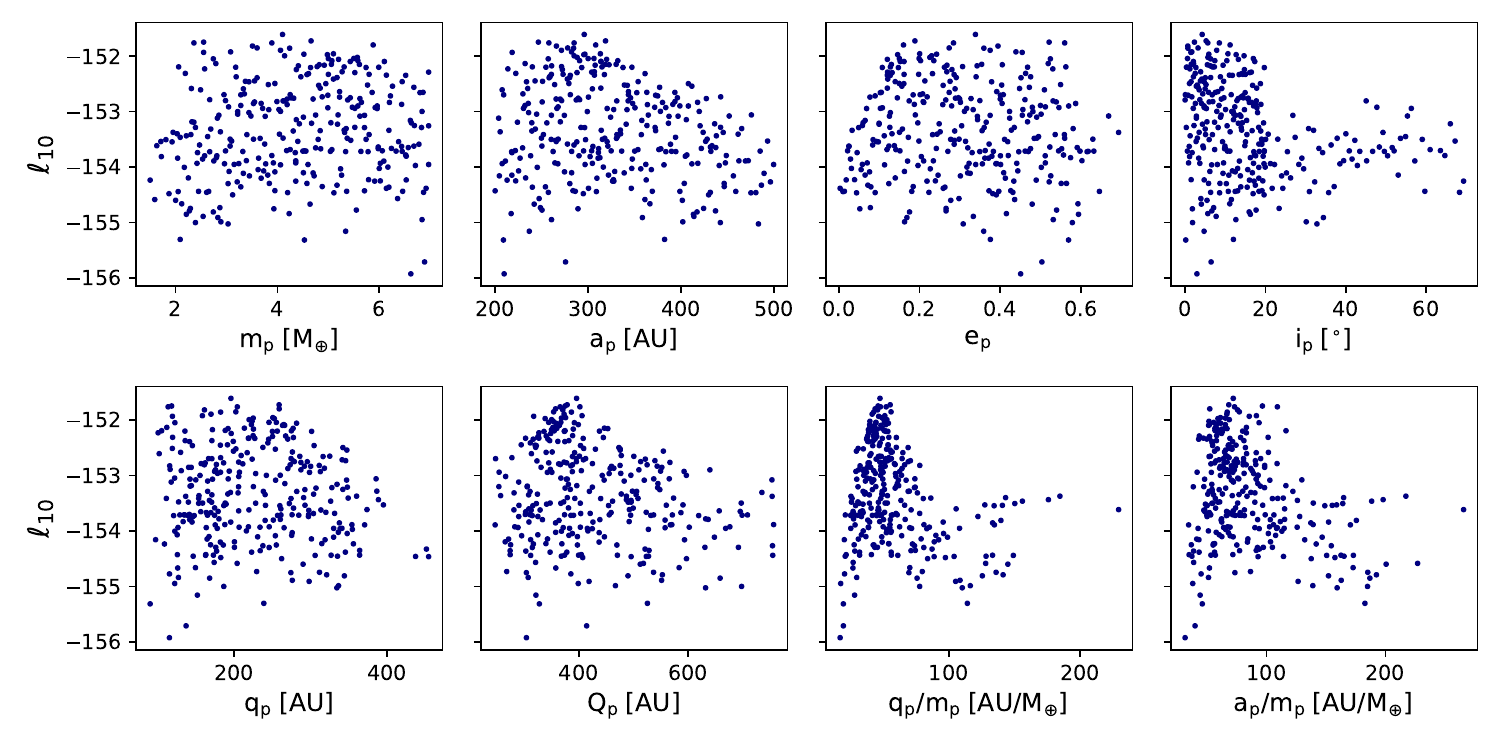}
\caption{Log likelihood, $\ell_{10}$, as a function of various orbital parameters for the 300 simulations.}
\label{fig:scatterplots}
\end{figure*}

\begin{figure}%[hptb]
\centering
\includegraphics[width=\linewidth]{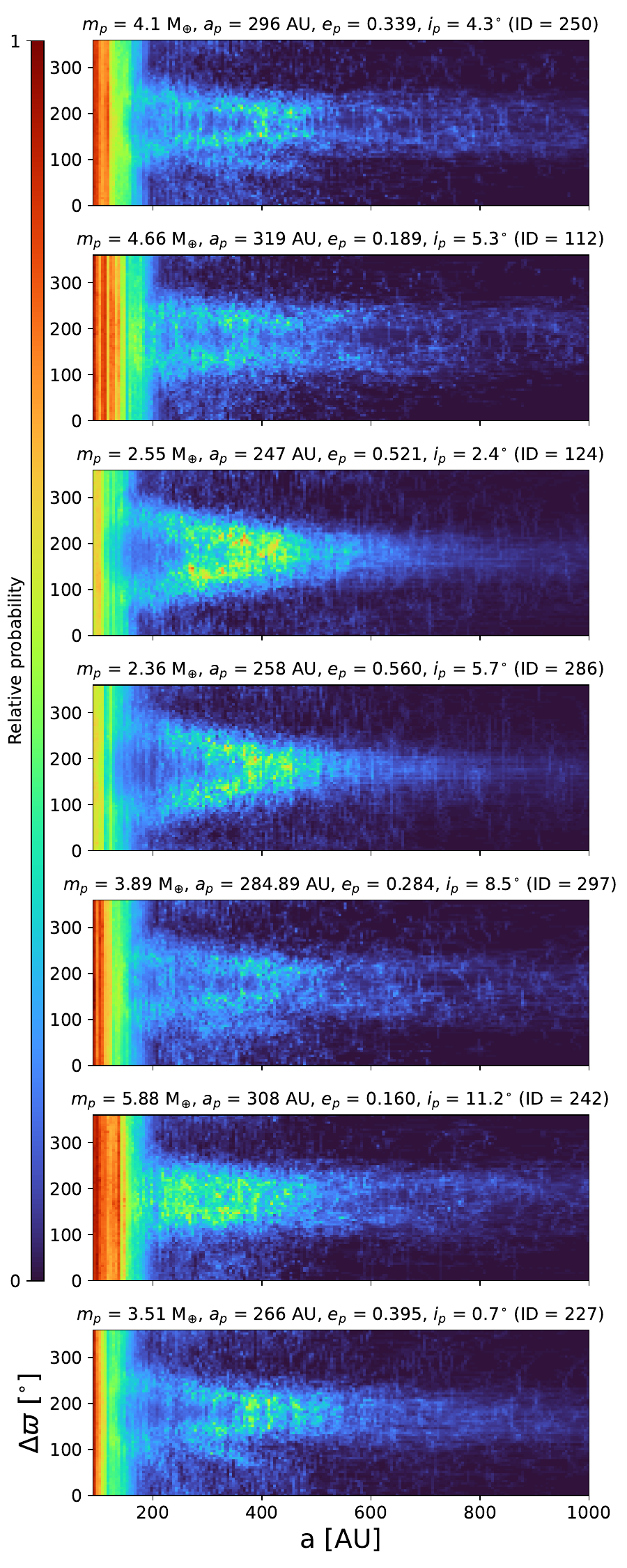}
\caption{Density maps of $\Delta \varpi$ as a function of $a$ for test particles with $35 \mathrm{\; AU} < q < 80 \mathrm{\; AU}$, corresponding to the 7 simulations that produced the highest values of $\ell_{10}$ (the top $1\sigma$ set).}
\label{fig:densitymaps}
\end{figure}

Each simulation contains 12\,960 test particles. The  values of semimajor axis, perihelion, and inclination are randomly selected from uniform distributions in the respective ranges of $70$--$500\mathrm{\; AU}$, $30$--$50 \mathrm{\; AU}$, and 0--$25^{\circ}$. For each test particle as well as for the additional planet, the longitude of perihelion, longitude of the ascending node, and mean anomaly are all randomly selected from uniform distributions in the range 0--$2\pi$. The other parameters of the hypothetical planet are mass $m_p$, semimajor axis $a_p$, eccentricity $e_p$, and inclination $i_p$. 

We first integrate each system forward in time by $1 \mathrm{\; Gyr}$ so that the test particles reach a steady state \citep{2021AJ....162..219B}. After the initial $1 \mathrm{\; \mathrm{Gyr}}$ integration has completed, we record the orbital parameters of each particle in snapshots of the simulation at intervals of $20 {\; \mathrm{kyr}}$. Each simulation runs for an additional $\sim 1 \mathrm{\; Gyr}$; particles that collide with the Sun or any of the planets are removed, as are particles with heliocentric distances that exceed $10^5 \mathrm{\; AU}$ which are assumed to have escaped. 
Each simulation produces a total of $\sim 5$--$6 \times 10^8$ recorded sets of test-particle orbital elements.  In total, we ran 300 simulations in four sets, described below.

To pick the parameters of the hypothetical planet for the first set of simulations, we employed latin hypercube sampling over the following ranges for $m_p$, $a_p$, $e_p$, and $\sin i_p$, respectively: $2$--$7 M_{\oplus}$, $200$--$500 \mathrm{\; AU}$, $0$--$0.7$, and $0$--$0.94$. We selected 98 such sets of parameters (\texttt{sim ID 1}--\texttt{sim ID 98}) that satisfy $q_p > 115 \mathrm{\; AU}$ to ensure that resonant TNOs in the 4:1 mean-motion resonance are preserved \citep[][see the end of this Section for further discussion]{2023AJ....166..118L}. Additionally, we ran the \cite{2021AJ....162..219B} preferred parameters ($m_p = 6.2 \; \mathrm{M_\oplus}$, $a_p = 380 \; \mathrm{AU}$, $e_p = 0.21$, $i_p = 16^{\circ}$) as \texttt{sim ID 99} and their simulation that yielded the highest likelihood ($m_p = 5 \; \mathrm{M_\oplus}$, $a_p = 300 \; \mathrm{AU}$, $e_p = 0.15$, $i_p = 17^{\circ}$) as \texttt{sim ID 100}. 

After each simulation run, we evaluate the fit to the observed distribution of $\varpi$ as a function of semimajor axis for distant TNOs, which we show to be statistically significant and unaffected by observational selection effects in Sections \ref{kuipertestsection}, \ref{bias}, and \ref{simbias}, using the following procedure. The set of observed  stable/metastable TNOs\footnote{The distributions of test particles in each simulation are compared to the actual distributions of stable/metastable TNOs because by construction, the relevant test particles have remained stable over $1$--$2$ Gyr.} with $q > 30 \mathrm{\; AU}$ and $90 \mathrm{\; AU} < a < 1000 \mathrm{\; AU}$ contains objects with perihelia in the range $\sim 35$--$80 \mathrm{\; AU}$, so we first select all test particle snapshot orbits with $q$ in this range. From this perihelion-limited sample of orbits, we select the subset with semimajor axes similar to ($\pm 10\%$) that of each object in the observed population. For each subset, we construct a Gaussian kernel density estimate for the distribution of $\Delta\varpi \equiv \varpi - \varpi_p$ and evaluate the probability of producing the observed value of $\Delta\varpi$ for the respective object. For the simulated orbits, $\Delta\varpi$ is trivial to calculate because the longitude of perihelion for the planet, $\varpi_p$, is always known, but for the observed TNOs, a value of $\varpi_p$ must be adopted. We blindly search all possible values of $\varpi_p$ (0--$2 \pi$). For any choice of $\varpi_p$, we compute the base-10 log of the product of the probabilities for drawing each value of $\Delta\varpi$ from the respective subset of simulation orbits. We record the value of $\varpi_p$ that maximizes the log likelihood, as well as the log likelihood value itself, $\ell_{10}$.

Based on the simulations in the first set that yielded the highest log likelihood values, we defined a smaller region of parameter space to sample from for the second set of simulations. We selected 100 models (\texttt{sim ID 101}--\texttt{sim ID 200}) via latin hypercube sampling over the following ranges for $m_p$, $a_p$, $e_p$, and $i_p$, respectively: $1.5$--$7 M_{\oplus}$, $200$--$450 \mathrm{\; AU}$, $0.05$--$0.6$, and $0$--$20^{\circ}$. The most significant change from the first set of simulations was narrowing the range of inclinations considered, as models with $i_p \gtrsim 20^{\circ}$ produced low-likelihood values (see Fig.\ \ref{fig:scatterplots}). The ranges for $a_p$ and $e_p$ were narrowed slightly, and the lower bound on $m_p$ was decreased to ensure that a firm lower bound for a possible additional planet can be established. Additionally, we placed no constraint on $q_p$ \textit{a priori} for the second set of simulations.

Based on the planet parameters for the 10 simulations that produced the largest values of $\ell_{10}$ among the 200 simulations contained in the first and second sets, we defined two regions of parameter space to run the third set of simulations: one focused on low inclination models ($0^{\circ} < i < 7^{\circ}$) the other on moderate inclination models ($13^{\circ} < i < 20^{\circ}$). The low inclination models were selected using latin hypercube sampling over the following ranges for $m_p$, $a_p$, $e_p$, and $i_p$, respectively: $2$--$6 M_{\oplus}$, $200$--$350 \mathrm{\; AU}$, $0.1$--$0.55$, and $0$--$7^{\circ}$. The moderate inclination models were also selected using latin hypercube sampling, over the ranges of $4$--$6 M_{\oplus}$, $250$--$450 \mathrm{\; AU}$, $0.1$--$0.35$, and $13$--$20^{\circ}$, respectively. The relative number of simulations selected in the two phase space regions were chosen to be proportional to the respective phase-space volumes. We ran 30 additional low inclination simulations (\texttt{sim ID 201}--\texttt{sim ID 230}) and 11 additional moderate inclination simulations (\texttt{sim ID 231}--\texttt{sim ID 241}).

While no single values of $m_p$, $a_p$, $e_p$, or $i_p$ maximize $\ell_{10}$ across the 241 members of the first three sets of simulations, the simulations that produced the highest values of $\ell_{10}$ exhibit notably narrow ranges in aphelion distance, $Q_p \equiv a_p(1 + e_p)$, and perihelion distance divided by mass, $q_p / m_p \equiv a (1 - e_p) / m_p$ (see Fig.\ \ref{fig:scatterplots}). Therefore we used the fourth and final set of 59 simulations (\texttt{sim ID 242}--\texttt{sim ID 300}) to explore this parameter space in more detail. Specifically, we used latin hypercube sampling over the following ranges for $Q_p$, $q_p / m_p$, $m_p$, and $i_p$, respectively: $310$--$410 \mathrm{\; AU}$, $37$--$57 \mathrm{\; AU / M_{\oplus}}$, $2$--$7 M_{\oplus}$, and $0$--$20^{\circ}$. The ranges in $Q_p$ and $q_p / m_p$ cover $\sim 70\%$ and $\sim 80\%$ of the respective ranges for simulations that produced the top $2 \sigma$ of $\ell_{10}$ values\footnote{The top $n \sigma$ set of $\ell_{10}$ values is defined as those within $\log_{10}(e^{-n^2 / 2})$ of the highest value of $\ell_{10}$.} among the first three sets of simulations, while the ranges for $m_p$ and $i_p$ exceed the respective ranges for the top $2 \sigma$ of $\ell_{10}$ values in these simulations. 

The  parameters and resulting $\ell_{10}$ values for the top $2\sigma$ of the 300 simulations are listed in Table \ref{simulatedplanets}. Figure \ref{fig:scatterplots} shows how $\ell_{10}$ behaves as a function of various orbital parameters, and Figure \ref{fig:densitymaps} shows the $\Delta \varpi$ distribution as a function of $a$ for the simulations that produced the top $1 \sigma$ set of $\ell_{10}$ values. In addition to the 300 simulations, we also ran one simulation without an additional planet as a control –– this yielded $\ell_{10} = -154.62$, a value lower than $\sim 90\%$ of the simulations we ran that included an unseen planet.

\begin{figure*}%[hptb]
 \centering
\includegraphics[width=1\linewidth]{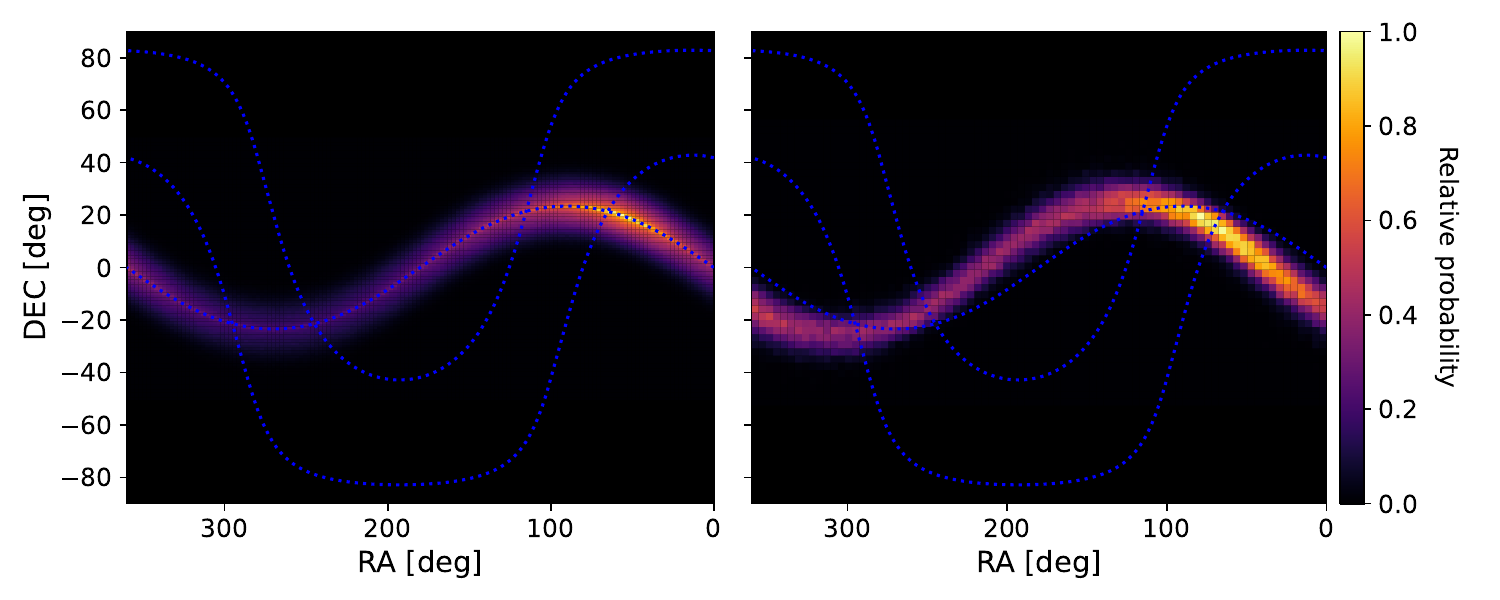}
\caption{Left: the relative probability density distribution for the unseen planet as a function of right ascension and declination. Right: the relative probability density distribution from \cite{2021AJ....162..219B}. Blue dashed lines indicate the ecliptic plane and $\pm 20^{\circ}$ from the Galactic plane.}
\label{fig:sims}
\end{figure*}
Any proposed additional planet in the solar system should be consistent with the long-term stability of TNO populations in mean-motion resonance (MMR) with Neptune \citep{2023AJ....166..118L, 2024PSJ.....5...61P}. The most distant populated $n:1$ resonances with $> 2$ members are the 4:1 and 5:1 resonances, which each contain 6 objects \citep{2024RNAAS...8...36V}. We first check which of these objects are long-term stable. We integrate the giant planets, plus 100 clones of each of the 12 resonant objects, for $4 \mathrm{\; Gyr}$. For only 4 of the 12 objects do $> 95 \%$ of the clones survive. All four are 4:1 MMR TNOs: 2003 LA7, 2013 SF106, 2015 RB278, and 2015 VO166. We use these objects to test the physical plausibility of various additional planet models. Specifically, for each model, we integrate for $4 \mathrm{\; Gyr}$ the giant planets, the additional planet, and 100 clones of each of the 4 highly stable 4:1 MMR TNOs. Clones are generated in the same way as described in Section \ref{longtermstability}. For the additional planet, the longitude of perihelion and the mean anomaly are chosen randomly from a uniform distribution over the range $0$--$2 \pi$. We consider any models for which fewer than half of all clones remained within the range of semimajor axis that would allow them to be in resonance with Neptune to be unphysical \citep{2019CeMDA.131...39L}. We find that four models are excluded on this basis (\texttt{sim ID 14}, \texttt{195}, \texttt{217}, and \texttt{251}), namely the four models with perihelion $q_p < 104 \mathrm{\; AU}$.

\section{Results}
\label{results}

\begin{figure}%[hptb]
 \centering
\includegraphics[width=\linewidth]{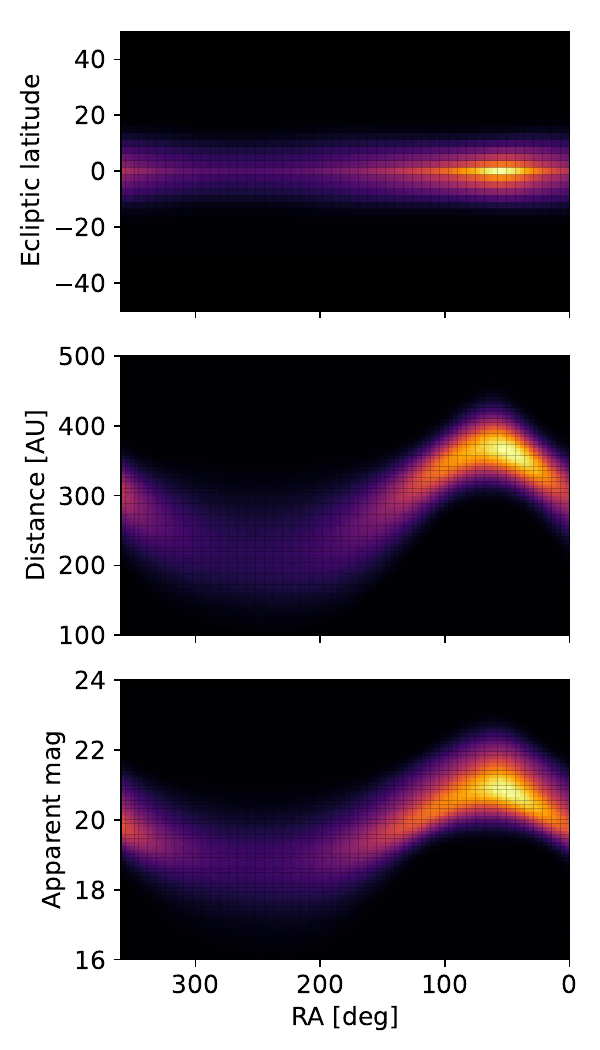}
\caption{Probability density for the unseen planet as a function of right ascension and, from top to bottom, ecliptic latitude, heliocentric distance, and apparent magnitude at opposition.}
\label{fig:sims_several}
\end{figure}

The simulation that produced the highest value of $\ell_{10}$ had the following unseen planet parameters: $m_p = 4.1 \mathrm{\; M_{\oplus}}$, $a_p = 296 \mathrm{\; AU}$, $e_p = 0.34$, and $i_p = 4.27^{\circ}$. The resulting $\ell_{10}=-151.61$ is $3.7 \sigma$ higher than the value produced by the simulation lacking an unseen planet. We verified the repeatability of our methodology by initializing a new simulation with the same parameters that produced the top-ranked simulation but with a different random seed, finding that the resulting value of $\ell_{10}$ remained the highest among all 300. 

The set of simulations that produce the top $1\sigma$ of $\ell_{10}$ values have narrow ranges in various quantities including $i_p \; (0.7\mbox{--}11.2^{\circ})$, $a_p \; (247\mbox{--}319 \mathrm{\; AU})$, $Q_p \; (358\mbox{--}402 \mathrm{\; AU})$, and $q_p/m_p \; (44\mbox{--}56 \mathrm{\; AU / M_{\oplus}})$, suggesting that the distribution is unimodal. Given the apparent strong covariances between parameters, we produce a continuous function for $\ell_{10}$ by fitting a quadrivariate Gaussian distribution to the results of the set of 58 simulations that produced the top $2 \sigma$ set of $\ell_{10}$ values among the 300, to avoid bias induced by simulations that produce poor fits to the data. If $\ell_{10,\rm pred}$ is the log likelihood predicted for a given set of planetary parameters by the Gaussian distribution, we may define a $\mathcal{X}^2$ statistic, $\langle (\ell_{10} - \ell_{10,\rm pred})^2 \rangle$ for the set of 58 simulations that produced the top $2 \sigma$ set of $\ell_{10}$ values. This statistic is $0.39$ (or $0.28$ if the largest outlier among the 58 is removed), indicating that the log likelihoods are reasonably well fit by a Gaussian.

The mean and standard deviation for each of the four parameters involved in the quadrivariate Gaussian distribution are $m_p = 4.4 \pm 1.1 \mathrm{\; M_{\oplus}}$, $a_p = 290 \pm 30 \mathrm{\; AU}$, $e_p = 0.29 \pm 0.13$, $i_p = 6.8 \pm 5.0^{\circ}$. Due to strong covariances between the parameters, certain combinations of the parameters are particularly well-confined, such as aphelion, $Q_p = 370 \pm 30 \mathrm{\; AU}$, and the perihelion-to-mass ratio, $q_p / m_p = 46.8^{+6.1}_{-5.1} \mathrm{\; AU \; / \; M_{\oplus}}$. The full covariance matrix is

\[
  \left[ {\begin{array}{cccc}
   1.3 & 26 & -0.13 & 2.3 \\
   26 & 890 & -2.8 & 34 \\
   -0.13 & -2.8 & 0.018 & -0.33 \\
   2.3 & 34 & -0.33 & 25 \\
  \end{array} } \right] \; \; ,
\]
where the rows and columns are in the order, $m_p$, $a_p$, $e_p$, $i_p$. 

We note that the aforementioned covariance matrix for $m_p$, $a_p$, $e_p$, and $i_p$ can be diagonalized for the following linear transformation of the parameters,
\begin{align*}
a_p & = a_p \; \; , \\
b_p & = \frac{i_p}{\mathrm{deg}} - \frac{a_p}{\mathrm{25 \; AU}} \; \; ,\\ 
c_p & = \frac{m_p}{\mathrm{M_{\oplus}}} - \frac{a_p}{\mathrm{25 \; AU}} - \frac{i_p}{\mathrm{17^{\circ}}} \; \; ,\\
d_p & = e_p + \frac{m_p}{\mathrm{12 \; M_{\oplus}}} + \frac{i_p}{\mathrm{200^{\circ}}} + \frac{a_p}{\mathrm{2400 \; AU}} \; \; .
\end{align*}
The mean and standard deviation for $b_p$, $c_p$, and $d_p$ are as follows: $b_p = -4.7 \pm 5.0$, $c_p = -3.8 \pm 0.7$, $d_p = 0.81 \pm 0.06$.

To derive the probability distributions for the unseen planet's heliocentric distance, right ascension and declination, we employ a Monte Carlo method. We sample $m_p$, $a_p$, $e_p$, and $i_p$ from the quadrivariate Gaussian distribution described above, $\varpi_p$ from a uniform distribution in the range $238^{\circ}$--$254^{\circ}$ (the range spanned by the top $2 \sigma$ simulations), and $\Omega_p$ and mean anomaly $M_p$ from a uniform distribution in the range $0$--$2 \pi$. We require $m_p > 0$, $a_p > 0$, $0 \leq e_p \leq 1$, $i_p \geq 0$, and $q_p > 104 \mathrm{\; AU}$ (see discussion at the end of Section \ref{sec:simul}). Taken together, the six Keplerian orbital elements allow us to calculate heliocentric distance, right ascension, and declination. To estimate the apparent magnitude, we ignore the difference between heliocentric and geocentric distance. We adopt an albedo from a uniform distribution in the range $0.1$--$0.6$ and a radius derived from the mass-radius relation described in \cite{2024A&A...686A.296M}, although we note that an unseen planet in the solar system would have a considerably lower temperature than the known exoplanets used to construct mass-radius relations. For the known exoplanet population, \cite{2024A&A...686A.296M} argue that any planet with $M < 4.4 M_{\earth}$ is most likely rocky; we would expect the mass-radius relation for such worlds to have little temperature dependence. We repeat this procedure a large number of times, the result of which is displayed in Figures \ref{fig:sims} and \ref{fig:sims_several}. Our simulated population is clustered around the ecliptic plane, with $90\%$ of planets within $\pm 10^{\circ}$ of the ecliptic. The majority of planets ($60\%$) are in the Northern sky but have an ecliptic latitude less than $+10^{\circ}$, a region of the sky that will be surveyed by the Vera C.\ Rubin Observatory's Legacy Survey of Space and Time (LSST), primarily through the North Ecliptic Spur (NES) mini-survey \citep{2018arXiv181201149S}. A similar fraction of planets ($60\%$) are expected to have a visual magnitude $V > 20$, which would elude many existing, pre-LSST efforts. The probability density peak in right ascension occurs at $\sim 55^{\circ}$ and half of all planets have right ascension in the range $0 - 110^{\circ}$. Interestingly, all planets are brighter than the LSST flux limit.

We find that only $0.06\%$ of the \cite{2021AJ....162..219B} reference population \citep{brown_2023}\footnote{\url{https://data.caltech.edu/records/8fjad-x7y61}} produce probabilities within $1\sigma$ of the maximum within our quadrivariate Gaussian model. This is a factor of 150 lower than the fraction of a quadrivariate Gaussian's probability density function within $1\sigma$ of the maximum, indicating that this work has identified a different unseen planet candidate than \cite{2021AJ....162..219B} -- lower in mass ($4.4\pm 1.1M_\oplus$ versus $6.2^{+2.2}_{-1.3}M_\oplus$), semimajor axis ($290\pm 30\mbox{\;AU}$ versus $380^{+140}_{-80} \mbox{\;AU}$), perihelion ($200\pm 50\mbox{\;AU}$ versus $300^{+85}_{-60} \mbox{\;AU}$), and inclination ($i_p = 7 \pm 5^{\circ}$ versus $i_p = 16 \pm 5^{\circ}$). We note that the range of $\varpi_p$ for the our top-likelihood simulations ($238$--$254^{\circ}$ for the top $2\sigma$ set) is consistent with the \cite{2021AJ....162..219B} result of $248^{+16}_{-14}$$^{\circ}$ ($\pm 1 \sigma$).

If we limit ourselves to the \cite{2021AJ....162..219B} sample of 11 TNOs, we find that the maximum-likelihood simulation in this work (\texttt{sim ID 250}) and the maximum-likelihood simulation in \cite{2021AJ....162..219B} (\texttt{sim ID 100}) produce indistinguishable likelihoods ($\Delta \ell_{10} = 0.01$), illustrating the advantage of using our expanded sample of 51 TNOs. Note that the region of ($m_p$, $a_p$, $e_p$, $i_p$) parameter space in which \texttt{sim ID 250} resides was not explored by \cite{2021AJ....162..219B}.

\section{Discussion}
\label{discussion}

Examining the distribution of distant TNOs filtered by a stability requirement allowed us to conclude that there is evidence for clustering in $\varpi$ at the $\sim 3 \sigma$ level, while clustering in other angular orbital elements are not statistically significant in the current sample. Comparing the results of 300 numerical simulations of the solar system plus an unseen planet to the actual distribution of $\varpi$ for distant stable/metastable TNOs as a function of semimajor axis, we identified a distinct region of parameter space in which an unseen planet is most likely: $m_p = 4.4 \pm 1.1 \mathrm{\; M_{\oplus}}$,\footnote{Note that a portion of our high-mass parameter space is excluded by constraints from Cassini telemetry \citep{fienga2020}.} $a_p = 290 \pm 30 \mathrm{\; AU}$, $e_p = 0.29 \pm 0.13$, $i_p = 6.8 \pm 5.0^{\circ}$, with the covariance matrix listed in Section \ref{results}. The simulations that produced the top $2 \sigma$ of $\ell_{10}$ log likelihood values had best-fit $\varpi_p$ values in the range $238$--$254^{\circ}$ in ecliptic coordinates. While the predicted planet from this work is distinct in $(m_p, a_p, e_p, i_p)$ phase space from the one described by \cite{2021AJ....162..219B}, the range of $\varpi_p$ resulting from our highest likelihood simulations is consistent with, albeit narrower than their estimate of $248^{+16}_{-14}$$^{\circ}$ ($\pm 1 \sigma$). The planet favored by our work is also capable of producing other phenomena ascribed to an unseen planet, including high-inclination distant ($i > 50^{\circ}$, $a > 250\mathrm{\; AU}$) TNOs and retrograde Centaurs ($i > 90^{\circ}$, $a > 30\mathrm{\; AU}$, $q < 30\mathrm{\; AU}$) \citep{2019PhR...805....1B, 2019AJ....158...43K}, and low-inclination TNOs on Neptune-crossing orbits \citep{2024ApJ...966L...8B}.

Our results differ from those of \cite{2021AJ....162..219B} for a few reasons. Firstly, our stability requirement is based on $4 \mathrm{\; Gyr}$ numerical integrations, rather than a perihelion cutoff, allowing for a clearer picture of the relevant TNOs. Secondly, the minimum value of $a$ for the sample of distant TNOs adopted in this work is $60 \mathrm{\; AU}$ lower than in \cite{2021AJ....162..219B}, allowing us to both find and fit the strong onset of $\varpi$ clustering as a function of $a$. These two reasons, combined with the increase in known TNOs over the past three years, account for the fact that this work fits simulation output to a sample of 51 TNOs, as opposed to the 11 used in \cite{2021AJ....162..219B}. Thirdly, we fit only to $\varpi$, and not to $\Omega$ and $i$, as $\varpi$ was the only orbital element of the three for which we found statistically significant clustering. Finally, we performed 300 simulations with varied unseen planet parameters, rather than 121, and allowed for smaller planet masses. 

While it is still possible that the clustering in $\varpi$ is due to observational selection effects, the evidence that stable/metastable and unstable populations have different distributions in $\varpi$ at the $\sim 3 \sigma$ level but similar observational selection effects suggests otherwise. As expected for a real effect, the statistical significance of $\varpi$ clustering increases with sample size -- from the individual survey level, at which is difficult to ascertain whether or not there is clustering in $\varpi$, to a group of surveys combined (see Appendices \ref{bias} and \ref{simbias}), to the entire sample (see Sections \ref{kuipertestsection} and \ref{evector}). Analysis of $\varpi$ clustering across the entire sample was enabled by our survey-independent methodology of comparing the clustering as filtered by the stability requirement. The lack of statistically significant clustering in the observed $\Omega$ and $\omega$ distributions of distant TNOs is consistent with an unseen planet's orbital plane being closely aligned with the ecliptic plane. Such an alignment (low $i_p$) was found in this work by fitting the $\varpi_p$ distribution alone. As $i_p$ grows, the amplitude of the clustering in $\varpi$ declines while the amplitudes of the clustering in $\Omega$ and $\omega$ both grow -- these trends are apparent in, for instance, the set of simulations with $m_p$, $a_p$, and $e_p$ values within $10\%$ of the top-likelihood simulation (\texttt{sim ID 250}).

The simulations that produce the highest values of $\ell_{10}$ exhibit narrow ranges in aphelion ($Q_p$) and perihelion-to-mass ratio ($q_p / m_p$): the $\pm 1 \sigma$ ranges in our quadrivariate Gaussian model are $370 \pm 30 \mathrm{\; AU}$ and $47^{+6}_{-5} \mathrm{\; AU / M_{\oplus}}$, respectively. The narrow range in $Q_p$ may arise because the planet spends most of its time near aphelion, so its effect on the TNOs is determined more strongly by $Q_p$ than by the other orbital elements. The narrow range in $q_p/m_p$ may arise because decreasing $q_p$ and increasing $m_p$ both tend to increase the strength of the planetary perturbation, and therefore there is some optimal value of the ratio $q_p / m_p$ that explains the observed degree of $\varpi$ confinement. 

The formation mechanism for such a planet may involve scattering by the giant planets followed by orbit circularization due to torques in the Sun's birth cluster or dynamical friction in an extended gaseous disk \citep{2016ApJ...826...64B, 2019PhR...805....1B}. The formation mechanism may leave an additional imprint on the orbital distributions of distant TNOs. This topic merits further study but is beyond the scope of this work.

We note that it is possible, as \cite{2021ApJ...910L..20B} posit, that the distant TNOs exhibiting orbital clustering may have a mixture of origins, some from the Kuiper belt (low inclinations and small semimajor axes) and some from the inner Oort cloud (high inclinations and large semimajor axes). In this work, as in \cite{2021AJ....162..219B}, we assume the Kuiper belt is the dominant source for objects with $a < 1000 \mathrm{\; AU}$, but if the inner Oort cloud contributes an appreciable amount, the expected orbital properties of the unseen planet would differ. In such a case, the expected eccentricity of the planet may increase \citep{2021ApJ...910L..20B}, possibly making it harder to detect.

Another possible source of clustering in $\varpi$ is resonances between distant TNOs and Neptune. Consider a resonance with a critical or slow angle $\psi=p\lambda -\lambda_N-(p-1)\varpi$, where $\lambda$ and $\lambda_N$ are the mean longitudes of the TNO and Neptune. The TNO is likely to be discovered near perihelion, so $\lambda\simeq\varpi$ and $\psi\simeq \lambda-\lambda_N$. Thus if $\psi$ librates with small amplitude around $\psi_0$, $\lambda$ and hence $\varpi$ will be concentrated around $\psi_0+\lambda_N$, where $\lambda_N$ is approximately constant over the limited duration of modern TNO surveys. \cite{2022ApJ...937..119V} investigated this effect and found that Neptune's resonances impart only a modest (few percent) nonuniformity in the distribution of $\varpi$ for the currently observable distant TNOs, too small to explain the observed clustering. We have performed an additional check using the simulation from \S\ref{sec:sim} that lacks an unseen planet. As described in that section, we integrated 12\,960 test particles for 1 Gyr, which should be long enough for the effects of resonant sticking to be established. We then integrate the simulation for an additional $10 \mathrm{\; Myr}$, recording the value of the angle $\psi=\lambda - \lambda_N$ for each particle at the time of each perihelion passage. For each particle with $\mathrm{170 \; AU} < a < \mathrm{1000 \; AU}$ and $\mathrm{35 \; AU} < q < \mathrm{80 \; AU}$, we calculate $C = \langle \cos{\psi} \rangle$ and $S = \langle \sin{\psi} \rangle$, where the average $\langle\cdot\rangle$ is over all perihelion passages. Large values of $A^2=C^2+S^2$ imply that $\psi$ is approximately constant and hence that the particle is in resonance. In the overall population of test particles we find that the $\psi$ distribution is nearly uniform, with variations only at the level of $\sim 1 \%$. Assuming that running our test particle simulation for 1 Gyr populates the resonances at the same rate as the actual evolution of TNOs, this implies that distant resonances of Neptune do not significantly affect the observed $\varpi$ distribution of TNOs. The $\sim 3 \%$ of test particles with $A^2 \gtrsim 0.5$ show a strongly non-uniform distribution in $\psi$, suggesting that most of these are in resonance. However, this distribution has two peaks separated by $\sim 200^{\circ}$ -- consistent with the strongest resonances seen in the analysis of \cite{2022ApJ...937..119V} -- suggesting that even if most observed distant TNOs were in resonance with Neptune, the observed $\varpi$ distribution could not be easily explained since it has a single peak. We conclude that resonances alone do not explain the observed asymmetry in $\varpi$.

\begin{ThreePartTable}
%\renewcommand\TPTminimum{\textwidth}
%% Arrange for "longtable" to take up full width of text block
\begin{TableNotes}
  \item[] \textsc{Note --} $\ell_{10}$ is the base-10 log of the likelihood for each simulation, calculated as described in Section \ref{sec:simul}. The asterisk next to \texttt{sim ID 217} signifies that it is implausible because it is inconsistent with the survival of highly stable TNOs in mean-motion resonance with Neptune (see discussion at the end of Section \ref{sec:simul}). The units for $m_p$ and $a_p$, omitted from the header given space constraints, are $M_{\oplus}$ and $\mathrm{AU}$, respectively.
\end{TableNotes}

\begin{longtable}{llllllll}
\caption{Parameters of simulated unseen planets that produced the top $2\sigma$ subset of log likelihood ($\ell_{10}$) values.}\\ 
\hline
$m_p$ & $a_p$ & $e_p$ & $i_p [\mathrm{^{\circ}}]$ & $\varpi_p [\mathrm{^{\circ}}]$ & $\ell_{10}$ & $\Delta \ell_{10}$ & ID  \\ \hline
\endfirsthead

%\endhead
%\hline
\vspace{-0.55cm} 
\endfoot
 % tell LaTeX where to insert the table-related notes
\insertTableNotes
\endlastfoot

\label{simulatedplanets}
%%-
4.1   & 296   & 0.339 & 4.27  & 242        & $-$151.61     & 0                  & 250  \\
4.66  & 319   & 0.189 & 5.25  & 242        & $-$151.73     & $-$0.12              & 112  \\
2.55  & 247   & 0.521 & 2.35  & 249        & $-$151.75     & $-$0.14              & 124  \\
2.36  & 258   & 0.56  & 5.66  & 254        & $-$151.76     & $-$0.15              & 286  \\
3.89  & 285   & 0.284 & 8.53  & 249        & $-$151.76     & $-$0.15              & 297  \\
5.88  & 308   & 0.16  & 11.21 & 241        & $-$151.8      & $-$0.19              & 242  \\
3.51  & 266   & 0.394 & 0.7   & 245        & $-$151.82     & $-$0.21              & 227  \\
3.6   & 278   & 0.273 & 3.24  & 245        & $-$151.85     & $-$0.24              & 249  \\
4.24  & 284   & 0.359 & 0.72  & 244        & $-$151.86     & $-$0.25              & 292  \\
4.06  & 271   & 0.375 & 7.88  & 245        & $-$151.89     & $-$0.28              & 276  \\
3.08  & 282   & 0.439 & 1.69  & 250        & $-$151.92     & $-$0.31              & 258  \\
2.55  & 218   & 0.455 & 1.46  & 247        & $-$151.93     & $-$0.32              & 62   \\
4.52  & 297   & 0.242 & 5.92  & 244        & $-$151.96     & $-$0.36              & 221  \\
5.1   & 313   & 0.2   & 10.68 & 239        & $-$151.96     & $-$0.35              & 246  \\
4.98  & 295   & 0.146 & 12.71 & 242        & $-$151.97     & $-$0.36              & 285  \\
4.14  & 285   & 0.23  & 14.68 & 246        & $-$151.99     & $-$0.38              & 166  \\
5.57  & 288   & 0.225 & 5.7   & 242        & $-$152.02     & $-$0.41              & 230  \\
5     & 300   & 0.15  & 17    & 245        & $-$152.04     & $-$0.43              & 100   \\
5.5   & 306   & 0.199 & 0.62  & 242        & $-$152.04     & $-$0.43              & 185  \\
2.74  & 257   & 0.524 & 6.38  & 252        & $-$152.05     & $-$0.44              & 245  \\
5.2   & 320   & 0.12  & 13.27 & 249        & $-$152.06     & $-$0.45              & 240  \\
5.07  & 332   & 0.202 & 7.41  & 243        & $-$152.08     & $-$0.47              & 275  \\
5.58  & 291   & 0.235 & 10.4  & 242        & $-$152.08     & $-$0.47              & 295  \\
6.1   & 314   & 0.152 & 17.35 & 244        & $-$152.1      & $-$0.49              & 243  \\
5.43  & 314   & 0.162 & 2.08  & 238        & $-$152.1      & $-$0.49              & 278  \\
2.79  & 232   & 0.519 & 2.11  & 245        & $-$152.13     & $-$0.52              & 205  \\
5.61  & 318   & 0.13  & 1.89  & 242        & $-$152.15     & $-$0.54              & 201  \\
5.06  & 330   & 0.354 & 1.55  & 238        & $-$152.15     & $-$0.54              & 202  \\
4.82  & 322   & 0.405 & 5.56  & 244        & $-$152.15     & $-$0.55              & 218  \\
4.92  & 308   & 0.268 & 9.46  & 240        & $-$152.16     & $-$0.55              & 300  \\
4.41  & 261   & 0.343 & 6.3   & 246        & $-$152.17     & $-$0.57              & 280  \\
4.78  & 271   & 0.307 & 6.74  & 246        & $-$152.19     & $-$0.58              & 229  \\
2.06  & 240   & 0.563 & 7.3   & 254        & $-$152.19     & $-$0.58              & 290  \\
5.31  & 315   & 0.128 & 6.93  & 239        & $-$152.19     & $-$0.59              & 298  \\
3.18  & 255   & 0.468 & 0.36  & 244        & $-$152.2      & $-$0.59              & 162  \\
5.99  & 347   & 0.262 & 0.48  & 240        & $-$152.2      & $-$0.59              & 208  \\
5.74  & 338   & 0.108 & 2.71  & 244        & $-$152.2      & $-$0.6               & 212  \\
4.65  & 293   & 0.148 & 19.72 & 245        & $-$152.21     & $-$0.6               & 273  \\
2.57  & 214   & 0.53  & 1.21  & 243        & $-$152.23     & $-$0.62              & 217* \\
4.37  & 272   & 0.28  & 14.78 & 247        & $-$152.24     & $-$0.63              & 252  \\
5.27  & 307   & 0.268 & 3.88  & 238        & $-$152.28     & $-$0.67              & 211  \\
6.97  & 304   & 0.121 & 16.49 & 244        & $-$152.29     & $-$0.68              & 287  \\
5.52  & 288   & 0.132 & 16.1  & 242        & $-$152.3      & $-$0.69              & 255  \\
2.19  & 222   & 0.463 & 2.54  & 245        & $-$152.31     & $-$0.7               & 209  \\
4.61  & 361   & 0.232 & 14.09 & 246        & $-$152.32     & $-$0.71              & 237  \\
5.8   & 273   & 0.198 & 14.38 & 243        & $-$152.33     & $-$0.72              & 234  \\
5     & 260   & 0.162 & 15.02 & 247        & $-$152.33     & $-$0.72              & 239  \\
4.58  & 318   & 0.286 & 13.44 & 242        & $-$152.33     & $-$0.72              & 293  \\
6.14  & 307   & 0.12  & 11.05 & 246        & $-$152.34     & $-$0.73              & 284  \\
6.52  & 281   & 0.127 & 11.78 & 246        & $-$152.35     & $-$0.74              & 299  \\
4.46  & 259   & 0.34  & 2.1   & 243        & $-$152.37     & $-$0.76              & 174  \\
3.19  & 259   & 0.476 & 8.81  & 253        & $-$152.37     & $-$0.76              & 269  \\
5.17  & 280   & 0.291 & 6.38  & 247        & $-$152.39     & $-$0.78              & 210  \\
3.6   & 258   & 0.452 & 5.03  & 248        & $-$152.39     & $-$0.78              & 214  \\
5.1   & 278   & 0.121 & 4.19  & 239        & $-$152.41     & $-$0.8               & 149  \\
3.55  & 243   & 0.213 & 3.83  & 242        & $-$152.45     & $-$0.84              & 177  \\
3.37  & 251   & 0.276 & 9.62  & 249        & $-$152.45     & $-$0.84              & 259  \\
6.45  & 354   & 0.14  & 1.5   & 243        & $-$152.46     & $-$0.85              & 164  \\
3.44  & 234   & 0.379 & 4.75  & 242        & $-$152.46     & $-$0.85              & 247  \\ \hline
\end{longtable}

\end{ThreePartTable}

Nearly all of the parameter space for the unseen planet proposed here falls within LSST's field of view and detection limits, so if such a planet exists, it is likely to be discovered early on in the survey\footnote{This conclusion relies on our assumptions that the planet has albedo $>0.1$ and a specific mass-radius relation.}. LSST will simultaneously reveal whether the observed clustering of distant TNOs in $\varpi$ is real, an observational selection effect, or a statistical fluke, given the large number of expected TNO discoveries \citep{2009arXiv0912.0201L}.

\newpage
\section*{Acknowledgements}
We thank Brett Gladman, Caleb Lammers, Josh Winn, Konstantin Batygin, Mike Brown, Peter Melchior, and Ronan Hix for useful discussions. We are pleased to acknowledge that the work reported in this paper was substantially performed using the Princeton Research Computing resources at Princeton University which is consortium of groups led by the Princeton Institute for Computational Science and Engineering (PICSciE) and Office of Information Technology's Research Computing.

%\newpage

\appendix
\section{Survey-specific clustering analysis}

\subsection{Empirical bias}
\label{bias}

\begin{figure*}%[hptb]
 \centering
\includegraphics[width=\linewidth]{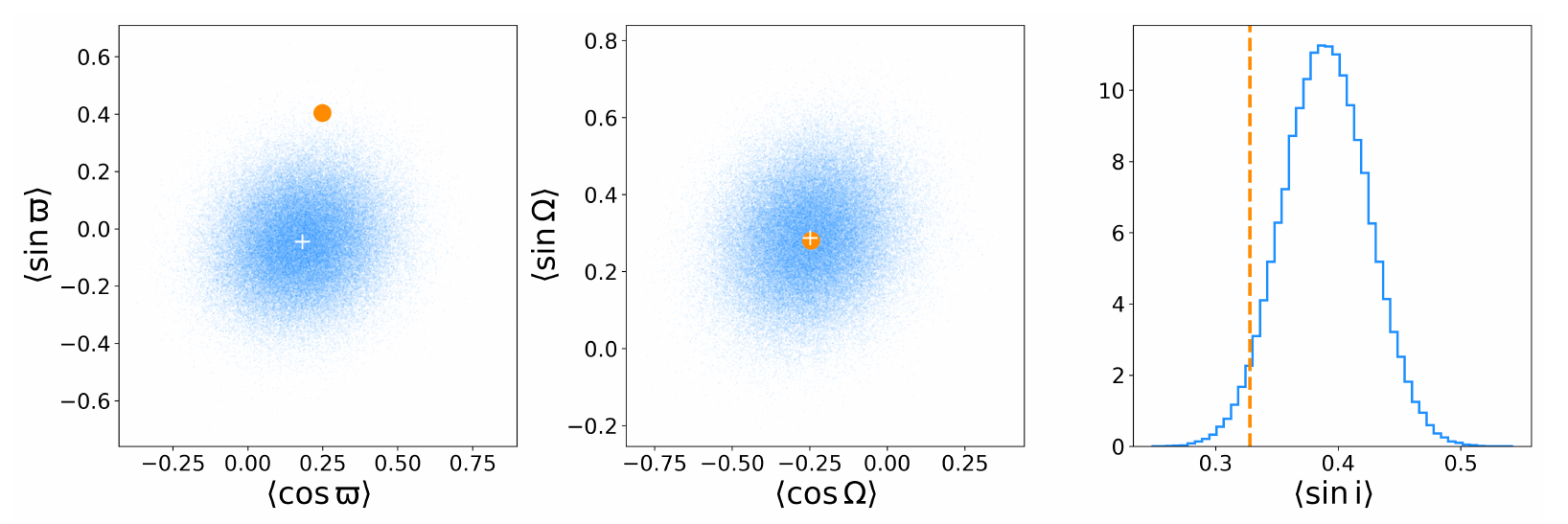}
\caption{Comparison of the mean elements of the 15 stable/metastable TNOs with $\mathrm{170 \; AU} < a < \mathrm{1000 \; AU}$ discovered by OSSOS, DES, and ST (in orange), with the means of $10^5$ synthetic samples of 15 objects constructed by sampling with replacement from discoveries made by the three surveys in the range $90 \mathrm{\; AU} < a < 170 \mathrm{\; AU}$.}
\label{fig:bias_plots}
\end{figure*}

Three surveys that have made major contributions to the catalog of distant TNOs are OSSOS \citep{2018ApJS..236...18B}, DES \citep{2020PSJ.....1...28B}, and Sheppard \& Trujillo (ST) \citep{2016AJ....152..221S}. Restricting the sample of TNOs to only those discovered by these three surveys provides an opportunity to study the possible influence of observational bias on apparent orbital clustering of TNOs.  In this light, we pose the question of whether or not distributions of orbital parameters for the population of stable/metastable objects with $\mathrm{170 \; AU} < a < \mathrm{1000 \; AU}$ differ at a statistically significant level from the distributions for TNOs discovered by the same surveys with $90 \mathrm{\; AU} < a < 170 \mathrm{\; AU}$, a semimajor axis range in which the experiments in Section \ref{kuipertestsection} showed that the distribution of $\varpi$ was uniform.

There are 15 stable/metastable TNOs with $\mathrm{170 \; AU} < a < \mathrm{1000 \; AU}$ discovered by OSSOS, DES, and ST (3, 6, and 6, respectively).\footnote{Only one of these objects was in the original \cite{2016AJ....151...22B} sample, which consisted of 6 objects. Of the remaining 5 objects from \cite{2016AJ....151...22B}, 3 are excluded because they were not seen by OSSOS, DES, or ST, and the other 2 are unstable. The sample of 11 TNOs used in \cite{2021AJ....162..219B} is entirely stable/metastable, but includes the same 3 objects from \cite{2016AJ....151...22B} that were not seen by OSSOS, DES, and ST in addition to a fourth such TNO, 2000 CR105. Of the 51 stable/metastable objects in our sample, 33 were discovered by OSSOS, DES, and ST.} To contextualize this observed sample, we generate synthetic samples of 15 objects from the population of TNOs discovered by DES, OSSOS, and ST with semimajor axes in the range $90 \mathrm{\; AU} < a < 170 \mathrm{\; AU}$ by sampling with replacement. For each sample, we allow the relative numbers of TNOs discovered by OSSOS, DES, and ST to vary by drawing from Poisson distributions characterized by the observed number of objects in the sample (3, 6, and 6, respectively), but we require the overall number of TNOs to be 15.

The left panel of Figure \ref{fig:bias_plots} show the values of $\langle\cos\varpi\rangle$ and $\langle\sin\varpi\rangle$, where $\langle\cdot\rangle$ denotes the average over the 15 TNOs in each sample for  $10^5$ synthetic samples constructed as described in the previous paragraph. Each ($\langle\cos\varpi\rangle$, $\langle\sin\varpi\rangle$) pair is displayed as a small blue point in \ref{fig:bias_plots}. The orange dot simply illustrates  ($\langle\cos\varpi\rangle$, $\langle\sin\varpi\rangle$) for the 15 stable/metastable TNOs with $170\mbox{\;AU} < a < 1000 \mbox{\;AU}$ found by OSSOS, DES, and ST. We find that only $0.8\%$ of the synthetic samples are more distant from the mean than the observed sample.  

The middle panel shows the same plot for the longitude of the node, $\Omega$. In contrast, in this case $99.8\%$ of the synthetic samples are further from the mean than the observed sample. The right panel shows the distribution of $\langle\sin i\rangle$, with the orange line serving the same function as the orange dot; here $8\%$ of the synthetic samples produce a value of $\langle\sin i \rangle$ that is further from the mean (on either side) than the observed sample with $170\mbox{\;AU} < a < 1000 \mbox{\;AU}$. 

We conclude that the distribution of $\Omega$ for the more distant ($\mathrm{170 \; AU} < a < \mathrm{1000 \; AU}$) stable/metastable TNOs discovered by the three surveys is completely consistent with the less distant ($\mathrm{90 \; AU} < a < \mathrm{170 \; AU}$) population of TNOs, and the difference observed in the distribution of $i$ is not statistically significant. As a result, in contrast to 
\cite{2021AJ....162..219B}, we do not use $\Omega$ or $i$ -- or the direction of the orbit normal, which is a combination of $\Omega$ and $i$ -- to interpret our simulations. The results of this analysis, which accounts for selection effects in the three largest surveys for distant TNOs, are consistent with the conclusions we reached in Section \ref{kuipertestsection}: statistically significant clustering is present in the longitude of perihelion $\varpi$ for semimajor axes $a>170\mbox{\;AU}$, but not in the other angular orbital elements.

\subsection{Simulated bias}
\label{simbias}

\begin{figure}%[hptb]
\centering
\includegraphics[width=0.5\linewidth]{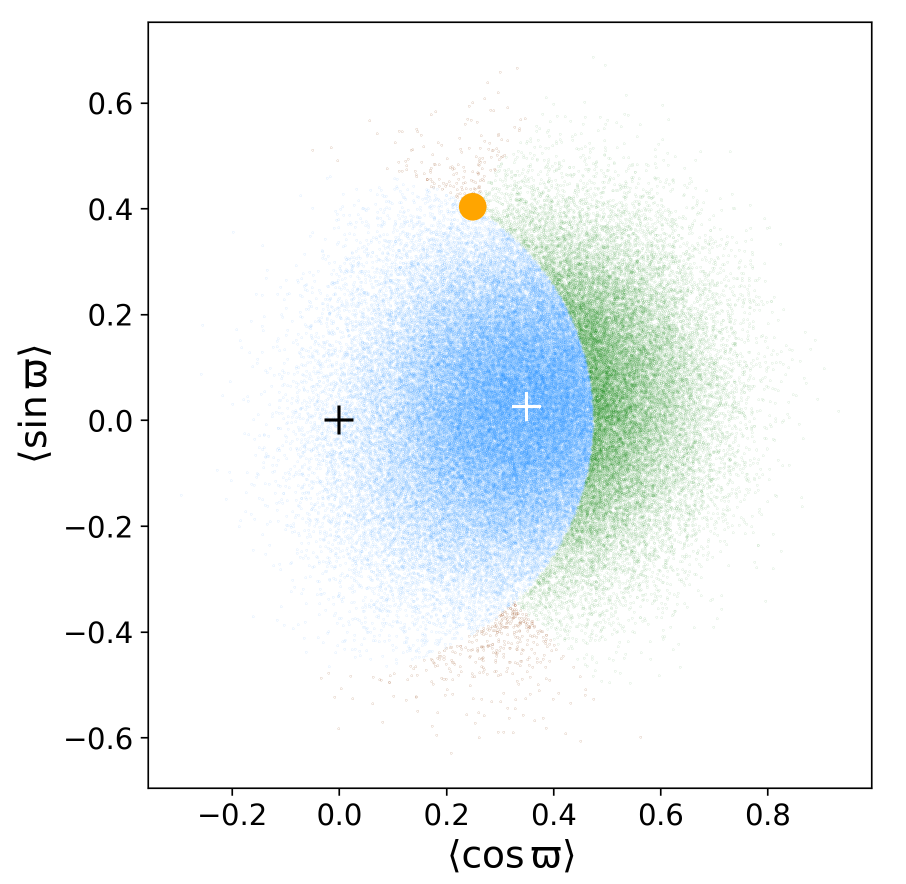}
\caption{Synthetic samples (small points in blue, green, and brown) versus the observed sample (large point in orange) of distant ($170 \mathrm{\; AU} < a < 1000 \mathrm{\; AU}$) stable/metastable TNOs discovered by OSSOS, DES, and ST. The coordinates are ($\langle\cos{\varpi}\rangle, \langle\sin{\varpi}\rangle$).  Synthetic samples were generated by drawing from the \cite{2021PSJ.....2...59N} simulated $q > 35 \mathrm{\; AU}$ detection efficiencies a uniformly distributed TNO population. The synthetic samples that produce clustering at least as strong as observed (i.e., radius from the origin larger than the radius of the orange dot) and at least as distant from the preferred direction as observed are highlighted in brown, while those that only produce clustering at least as strong as observed are highlighted in green. The rest of the synthetic samples are represented in blue. The white plus shows the average of all synthetic samples, and the black plus represents the origin.}
\label{fig:surveybias}
\end{figure}

In this Appendix we again ask whether the observed clustering in $\varpi$ in the OSSOS, DES and ST surveys implies that the underlying population of TNOs deviates from a uniform distribution, and is not simply a result of survey bias. This analysis is based on the $\varpi$ posteriors for simulated OSSOS, DES, and ST detections of TNOs distributed uniformly in $\varpi$, as computed by \cite{2021PSJ.....2...59N}.

We repeat the procedure described in Appendix \ref{bias}, but instead of drawing from the distribution of actual OSSOS, DES, and ST detections in the semimajor axis range $90 \mathrm{\; AU} < a < 170 \mathrm{\; AU}$, we drew from the \cite{2021PSJ.....2...59N} $\varpi$ posteriors for simulated detections of a uniformly distribution TNO population by the three surveys. Specifically, we used the $q > 35 \mathrm{\; AU}$ distribution in Figure 14 of \cite{2021PSJ.....2...59N}, as all of the objects in the observed sample satisfy $q > 35 \mathrm{\; AU}$. We drew $10^5$ samples of 15 $\varpi$ values, and computed the average values of $\cos \varpi$ and $\sin \varpi$ for each simulated synthetic sample. 

We find that there are strong survey selection effects evident in these samples of 15 TNOs, which bias them towards $\varpi \sim 6.6^{\circ}$. In contrast, the observed clustering for stable/metastable objects discovered by OSSOS, DES, and ST with semimajor axis in the range $170 \mathrm{\;AU} < a < 1000 \mathrm{\; AU}$ is towards $\varpi \sim 58.4^{\circ}$. To evaluate the likelihood that the observed clustering could be obtained from a population that is uniformly distributed in $\varpi$, we define the strength of clustering as the distance from the origin, $D\equiv \sqrt{\langle \cos{\varpi} \rangle^2 + \langle \sin{\varpi} \rangle^2}$, and the direction of clustering as $\Theta \equiv \arctan{\langle \sin{\varpi} \rangle / \langle \cos{\varpi} \rangle}$. We then ask what fraction 
of the synthetic samples produce clustering that is at least as strong as observed, $D>0.47$, and at least as distant from the value of $\varpi$ preferred by survey selection effects as observed, $|\Theta - 6.6^\circ| > 58.4^\circ-6.6^\circ$. The fraction of all synthetic samples that satisfy these constraints is only $0.5\%$.

We additionally check the likelihood that a group of 15 angles randomly chosen from the selection function for $\varpi$ (with the breakdown of objects from OSSOS, DES, and ST varying in the same way), when weighted by the inverse of the respective selection function, produces clustering as strong as observed in the actual sample (stable/metastable objects with semimajor axis in the range $170\mathrm{\; AU} < a < 1000 \mathrm{\; AU}$). We find that only $1\%$ of such groups produce average clustering at least as strong as observed. Both of these tests confirm that that survey bias is not a sufficient explanation for the observed distribution of $\varpi$ for the population of stable/metastable objects with semimajor axis in the range $170\mathrm{\;AU} < a < 1000 \mathrm{\; AU}$. This conclusion confirms the one reached by comparing the unstable population and the stable/metastable population with no cuts on discovery survey (see Figure \ref{fig:kuiper_statistic}).

%\newpage
\bibliography{bib}{}
\bibliographystyle{aasjournal}

\end{document}